# Orientation dependence of the Schottky barrier height for $La_{0.6}Sr_{0.4}MnO_3/SrTiO_3$ heterojunctions

M. Minohara[1], Y. Furukawa[2], R. Yasuhara[2], H. Kumigashira[2,3,4,a)], and M. Oshima[1,2,3,4]

[1]*Graduate School of Arts and Sciences, The University of Tokyo, Tokyo 153-8902, Japan*

[2]*Department of Applied Chemistry, The University of Tokyo, Tokyo 113-8656, Japan*

[3]*Core Research for Evolutional Science and Technology (CREST), Japan Science and Technology Agency, Tokyo 113-8656, Japan*

[4]*Synchrotron Radiation Research Organization, The University of Tokyo, Tokyo 113-8656, Japan*

**Abstract**

The authors report on the crystallographic orientation dependence of the Schottky properties for heterojunctions between a half-metallic ferromagnet $La_{0.6}Sr_{0.4}MnO_3$ (LSMO) and Nb-doped $SrTiO_3$ semiconductor. The Schottky barrier height determined by *in situ* photoemission measurements is independent for the substrate orientations (001) and (110), while the magnetic properties of LSMO (110) films are more enhanced than for (001) films. These results suggest that the performance of magnetic devices based on ferromagnetic manganite is improved by using (110)-oriented substrates.

a) Author to whom correspondence should be addressed; Electronic mail : kumigashira@sr.t.u-tokyo.ac.jp

Spin tunnel junctions employing ferromagnetic oxides have attracted much attention because of their potential applications to future spintronic devices such as the tunneling magnetoresistance (TMR) device [1,2]. The magnetic properties of ferromagnetic materials used as electrodes and tunnel barrier heights at ferromagnetic/insulator heterojunctions are essential parameters that dominate the performance of such spintronic devices. The ferromagnetic manganese oxide $La_{0.6}Sr_{0.4}MnO_3$ (LSMO) is a promising material for use in spintronic devices owing to its half-metallic ferromagnetic nature and high Curie temperature [3]. However, the performance of TMR devices based on a (001)-oriented LSMO/$SrTiO_3$ (STO)/LSMO structure is far worse than what would be expected from the high spin polarization of LSMO; this suggests the formation of an interfacial dead layer with degraded magnetic properties [1,2]. Recently, it has been reported that (110)-oriented films of optimally doped ferromagnetic manganese oxides present enhanced magnetic properties compared to (001)-oriented films, with a higher saturation magnetization and Curie temperature [4]. Furthermore, the thickness of the interfacial dead layer formed at the (110)-oriented $La_{2/3}Ca_{1/3}MnO_3$ (LCMO) /STO heterointerface is thinner than for the (001)-oriented one [5]. Thus, it is expected that the device performance should be improved by using a (110)-oriented spin tunneling heterojunction owing to the enhanced magnetic properties.

On the other hand, there have been few investigations on the tunnel barrier height for the (110)-oriented ferromagnetic metal/insulator heterojunction [6–12]. In particular, for the heterointerface of the (001)-oriented spin tunneling junction, the tunnel barrier height has been reported to be significantly higher than that predicted by the Schottky-Mott rule owing to the formation of the "interface dipole" at the interface [8,9]. To design spintronic devices with much higher performance, it is important to study the interfacial properties of (110)-oriented perovskite oxide heterojunctions.

In this study, we performed *in situ* photoemission spectroscopy (PES) characterization and current-voltage (*I-V*) measurements on (001)- and (110)-oriented LSMO/STO heterojunctions. The core-level PES measurement for buried STO layers enabled us to directly determine the barrier heights of the heterojunctions. The evaluated barrier heights of LSMO/STO (001) and (110) heterojunctions were $1.2 \pm 0.1$ and $1.1 \pm 0.1$ eV, respectively, indicating that the tunnel barrier height does not depend on the orientation. Through a comparison of the properties for the two orientations, we discuss the interfacial electronic structure of LSMO/STO heterojunctions.

LSMO thin films were grown on (001)- and (110)-oriented Nb-doped (0.05 wt%) STO (Nb:STO) substrates in a laser molecular beam epitaxy chamber connected to a synchrotron-radiation photoemission system at BL-2C of the Photon Factory in KEK [13]. Sintered LSMO pellets were used as targets. An Nd-doped yttrium aluminum garnet laser was used for ablation in its frequency-tripled mode ($\lambda = 355$ nm) at a repetition rate of 1 Hz. The (001)- and (110)-oriented Nb:STO substrates were annealed at 1050 ºC and an oxygen pressure of $1 \times 10^{-7}$ Torr to ensure an atomically flat surface [13,14]. During the *in situ* LSMO depositions, the substrate temperature and ambient oxygen pressure were 1000 ºC and $1 \times 10^{-4}$ Torr, respectively [13]. The film thicknesses were controlled on an atomic scale by monitoring the intensity oscillations of the reflection high-energy electron diffraction specular spot during growth. After deposition, the samples were moved into the photoemission chamber under a vacuum of $10^{-10}$ Torr. PES spectra were taken using a Scienta SES-100 electron energy analyzer with a total energy resolution of 150 meV in the 600–800 eV photon energy range.

The surface morphology of the films was analyzed using *ex situ* atomic force microscopy, which confirmed an atomically flat step-and-terrace structure. The crystal structures of epitaxial (001)- and (110)-oriented LSMO thin films were characterized through four-circle X-ray diffraction measurements. For the (110)-oriented LSMO film, a slight in-plane relaxation and practically full strain were observed along the [001] and [1−10] directions, respectively, while the (001)-oriented LSMO film was fully strained (not shown). Magnetization was measured using a superconducting quantum interference device magnetometer. Magnetic measurements were carried out with the magnetic field applied in-plane, with the (001) sample parallel to the [100] direction and the (110) sample parallel to the [100] and [110] directions. The temperature dependences of magnetization (*M-T*) for all films were recorded at 2 kOe, and magnetization hysteresis (*M-H*) loops were measured at 2 K with a maximum magnetic field of 5 kOe. *I-V* measurements were performed using the two-probe method at room temperature, which confirmed the Schottky contact for the (001)- and (110)-oriented LSMO/Nb:STO heterojunctions. In terms of the *I-V* characteristics, gold and aluminum electrodes—which are ohmic contacts for LSMO films and Nb:STO substrates, respectively, with diameters of 200 μm—were *ex situ* deposited on the films through thermal evaporation. A positive bias was defined as the current flow from the Nb:STO substrates to the LSMO films.

The *M-T* curves for the (001)- and (110)-oriented LSMO films with thicknesses of about 30 nm are shown in Fig. 1. The inset of Fig. 1 shows the *M-H* curves for both films. The *M-H* curves reveal the in-plane magnetic anisotropy of the (110)-oriented LSMO film due to its own crystallographic anisotropy. The magnetic anisotropic behavior of an LSMO (110) film agrees well with previous reports on (110)-oriented manganites [15, 16]. The Curie temperature of the (110)-oriented LSMO film is higher than for the (001)-oriented film by about 20 K. Infante *et al.* reported similar magnetic behavior for LCMO films and asserted that the origin of the enhanced magnetic properties of (110)-oriented perovskite manganite films was the absence of phase separation between ferromagnetic and nonferromagnetic regions, which were induced by in-plane relaxation of the films [4]. Since in-plane relaxation was also observed for the (110)-oriented LSMO film in this study, this strongly suggests that in-plane relaxation plays an important role in improving the magnetic properties of LSMO films.

Figure 2 shows the *I-V* characteristics for the (001)- and (110)-oriented LSMO/Nb:STO heterojunctions. Both of the heterojunctions exhibit the typical rectifying *I-V* characteristics for a Schottky junction composed of a metal and *n*-type semiconductor. Assuming that the transport is dominated by thermoionic emissions with a Richardson constant of 156 A cm$^{-2}$ K$^{-2}$ [7,17,18], the Schottky barrier heights (SBHs) are evaluated as 0.63 ± 0.02 and 0.55 ± 0.02 eV for the (001)- and (110)-oriented LSMO/Nb:STO heterojunctions, respectively. These values are in good agreement with the reported SBH determined from *I-V* measurements ($\Phi_B^{I-V}$): $\Phi_B^{I-V}$ for the (110)-oriented LSMO/Nb:STO heterojunction is 0.5–0.7 eV [6,7,9], while that of the (110) junction is 0.68 eV [12]. It was found that the difference in $\Phi_B^{I-V}$ between the (001) and (110) junctions is negligible. A similar orientation dependence was also observed for the Au/Nb:STO Schottky junction [19].

The negligibly weak dependence of SBHs on crystallographic orientation seems to be a common feature for metal/Nb:STO Schottky junctions. However, *I-V* measurements often underestimate the SBH owing to tunneling contributions; $\Phi_B^{I-V}$ is very sensitive to potential modulation in the interface region. Thus, in order to determine the real SBH (SBH at the edge of the depleted region), we performed Ti 2*p* core-level PES measurements on the two heterointerfaces; the band bending caused by the deposition of a metallic oxide film on Nb:STO can be directly determined from the Ti 2*p* core-level shift. The results are shown in Fig. 3. It should be noted that the measurements reflect the Ti 2*p* core-level position in the thin interface region only due to the short electron escape depth of 1–2 nm for the PES technique. For both junctions, a peak shift towards a lower binding energy was clearly observed as the overlayer film thickness increased. The energy shifts for the core-level peak position are summarized in Fig. 4. Judging from the saturation level of the peak shift, the energy shifts due to band bending can be estimated to be 1.2 ± 0.1 and 1.1 ± 0.1 eV for the (001) and (110) junctions, respectively. Since a flatband is formed at the surface of Nb:STO [8], these energy shifts directly correspond to the built-in potentials of the Schottky junctions. Owing to the negligibly small energy difference between the Fermi level and conduction band minimum (CBM) in the degenerate semiconductor Nb:STO [8], the SBHs determined by *in situ* PES ($\Phi_B^{PES}$) for the (001)- and (110)-oriented LSMO/Nb:STO junctions are 1.2 ± 0.1 and 1.1 ± 0.1 eV, respectively. The schematic band alignments for the (001)- and (110)-oriented LSMO/Nb:STO junctions, deduced from the present PES experiments, are illustrated in the inset of Fig. 4. The $\Phi_B^{PES}$ values are also independent from the orientation similar to $\Phi_B^{I-V}$, although there is a significant difference between $\Phi_B^{PES}$ and $\Phi_B^{I-V}$. These results suggest that the device performance may be improved by using (110)-oriented substrates owing to the enhanced magnetic properties of manganese films grown on (110) substrates.

When comparing the difference between $\Phi_B^{PES}$ and $\Phi_B^{I-V}$, $\Phi_B^{PES}$ is higher than $\Phi_B^{I-V}$ by 0.5–0.6 eV for both orientations. This difference may stem from the tunneling process in *I-V* measurements [11,20]. Assuming the existence of a thin depletion layer with an abrupt potential drop near the interface, the tunneling current dominates the *I-V* characteristics rather than the thermoionic emission current. Under such circumstances, the SBH is underestimated by *I-V* measurements, while $\Phi_B^{PES}$ can probe the real SBH at the edge of the depleted region because of the interface sensitivity of the PES measurements. The existence of a thin depletion layer at the interfaces may be in accordance with the significant broadenings of the Ti 2*p* core-level peak after LSMO deposition shown in Figs. 3 (a) and (b); the abrupt potential profile in the interface region comparable to the electron depth of the PES measurement makes the core level broaden [21,22].

In general, the narrowing of the depletion layer is caused by reduction of the permittivity and/or increase in donor concentrations. Although the origin of the narrowing of depletion layer near the interface is not clear at the moment, there are two possible explanations: one is the interfacial atomic rearrangement of constituent atoms, and the other is the modulation of stoichiometry in the interface region. When an LSMO film is deposited on Nb:STO, atomic rearrangement of constituent atoms, driven by the lattice mismatch between the two oxides, may occur to reduce the interface energy. As a result of this atomic rearrangement, the local permittivity of STO in the interface region is changed from that in bulk. Since STO is a quantum paraelectric material and its high permittivity originates from the delicate balance in its crystal structure, the atomic rearrangement at the interface may strongly influence the permittivity of STO in the interface region. On the other hand, by deposition of LSMO films on Nb:STO, oxygen vacancies in STO layers may be generated [23] and/or some interdiffusion may take place. The generation of oxygen vacancies and/or the occurrence of interdiffusion (La diffusion into STO layers) should cause an increase in the donor concentration in the interface region. In order to understand the origin of the difference between $\Phi_B^{PES}$ and $\Phi_B^{I-V}$, further systematic investigation is necessary. It is especially important to investigate the depth profiles of LSMO/Nb:STO junctions with PES for different escape depths.

In summary, we determined the Schottky barrier heights (SBHs) of (001)- and (110)-oriented LSMO/Nb:STO heterojunctions using *in situ* photoemission spectroscopy (PES) and *I-V* measurements. We found that SBHs do not depend on the orientation. In contrast, the magnetic properties of LSMO films grown on STO (110) substrates were more enhanced than for those on (001) substrates. These results suggest that the device performance based on perovskite manganites is improved through the use of (110)-oriented substrates. On the other hand, the significant difference in estimated SBHs between PES and *I-V* measurements suggests that a thin depletion layer with an abrupt potential drop exists in the interface region at the depth comparable to the probing depth for PES measurements.

We are very grateful to M. Lippmaa and M. Kawasaki for useful discussion and to I. Ohkubo and M. Kitamura for their support during the *I-V* measurements. Two of the authors (M. M. and R. Y.) acknowledge support from the Japan Society for the Promotion of Science (JSPS) for Young Scientists. This work was supported by a Grant-in-Aid for Scientific Research (S17101004 and A16204024) from the JSPS.

**Figure Captions**

FIG. 1. (color online): *M-T* curves of LSMO (001) (triangle) and (110) (circle) films of thickness 30 nm. The magnetic field is applied along the in-plane [100] direction for both films. The inset shows *M-H* curves for (001)- and (110)-oriented LSMO films.

FIG. 2. (color online): Typical |*I*|-*V* characteristics of (001)- and (110)-oriented LSMO/Nb:STO junctions (triangle and circle). The dashed straight lines are linear fits with the equation for the thermoionic emission process.

FIG. 3. (color online): Ti 2*p* core-level spectra of (a) Nb:STO (001) and (b) Nb:STO (110) covered by LSMO overlayers.

FIG. 4. (color online): Plot of the energy shift of the Ti 2*p* core-level peaks as a function of LSMO (001) (triangle) and (110) (circle) overlayer thicknesses. The lines are guides for the eye. The inset shows the band alignment deduced from present PES measurements of (001)- and (110)- oriented LSMO/Nb:STO junctions.

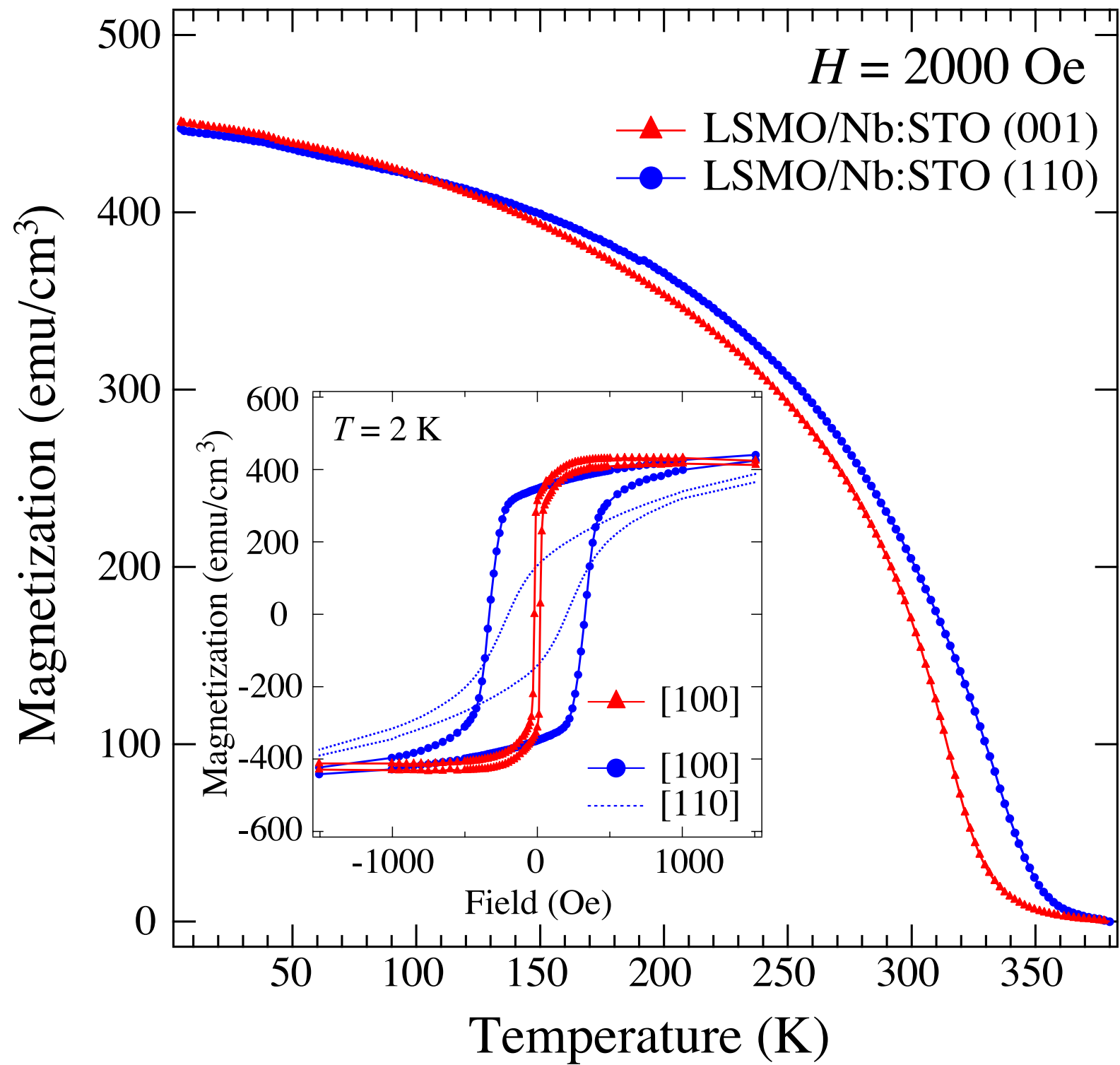

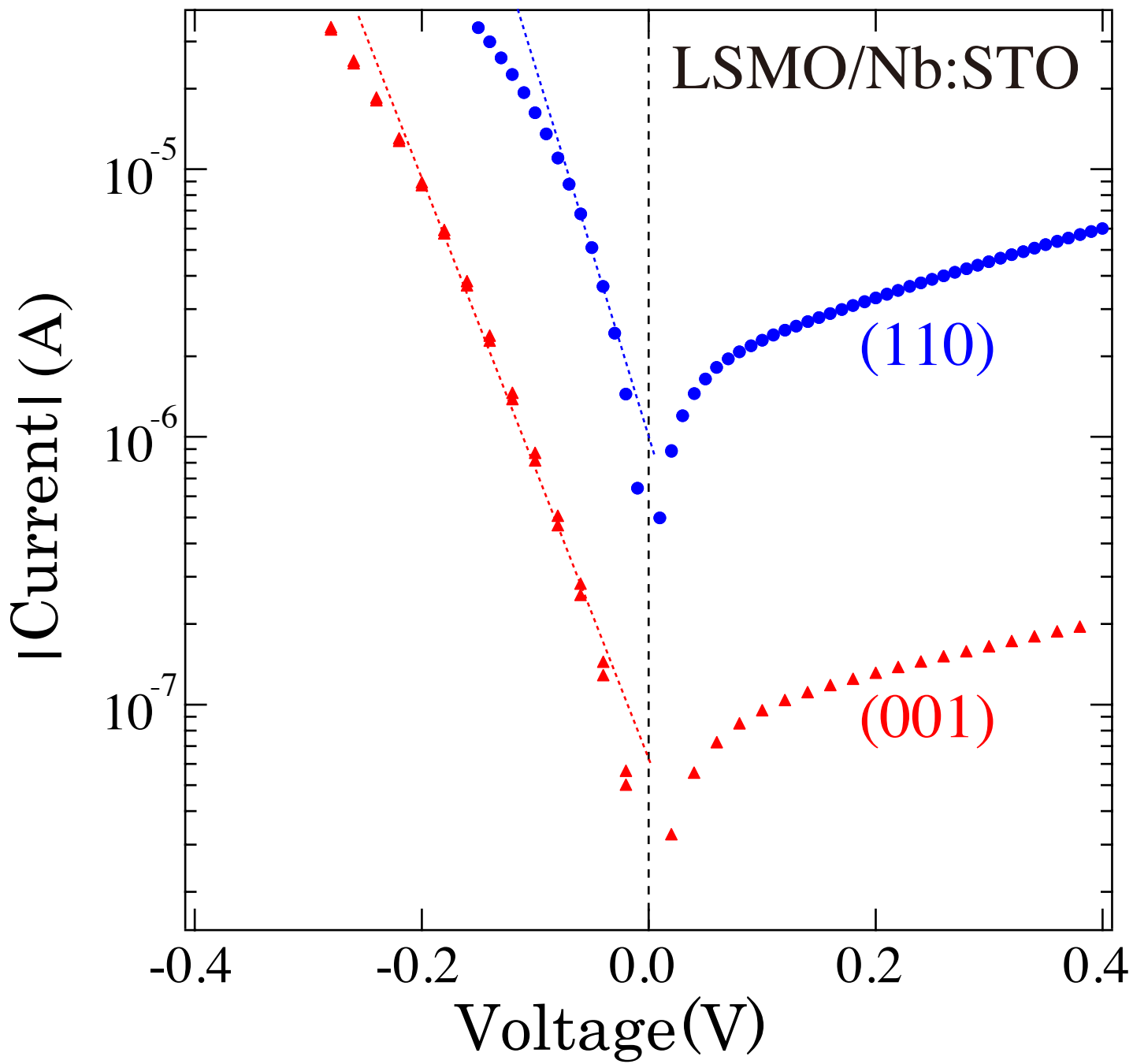

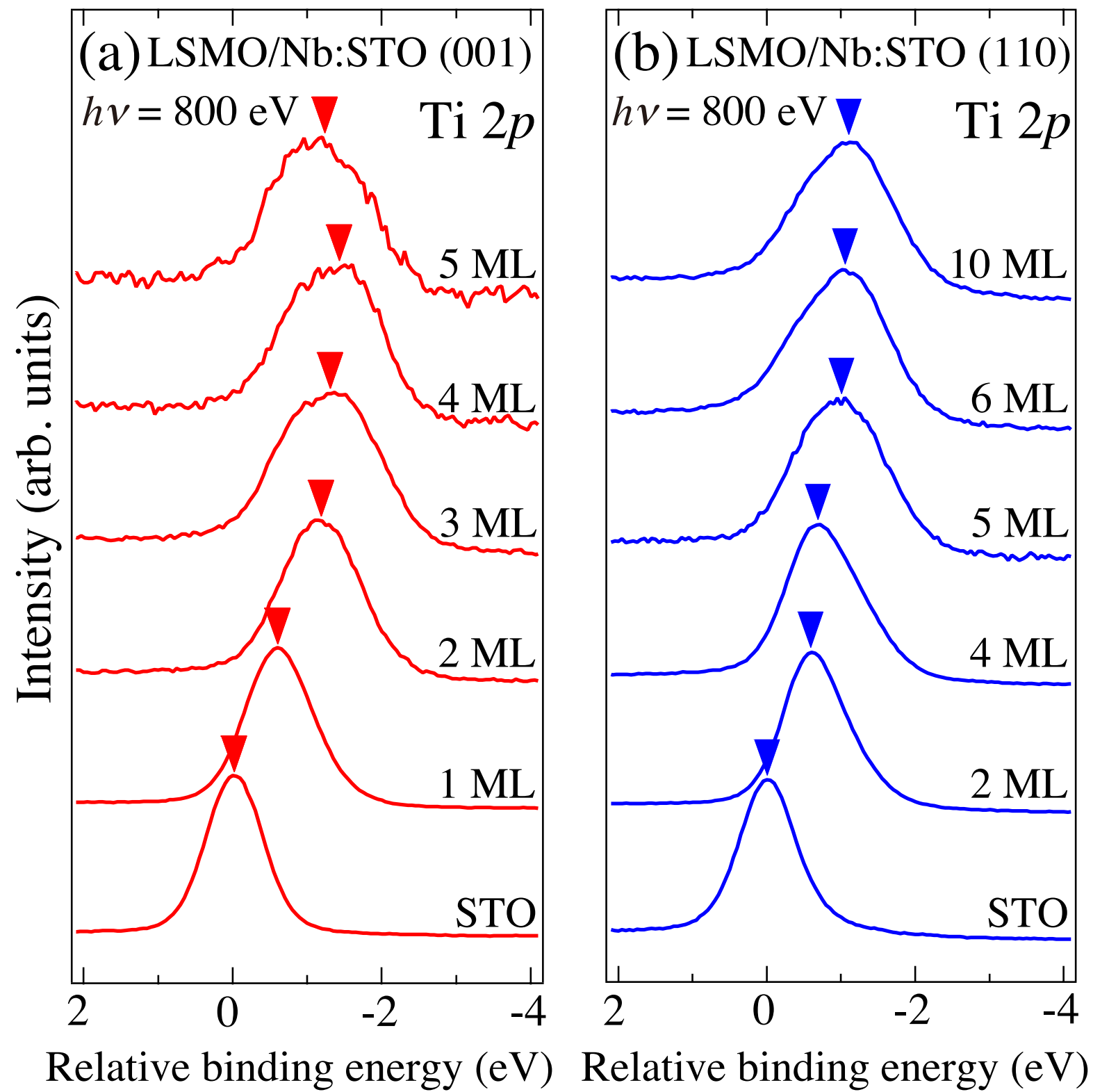

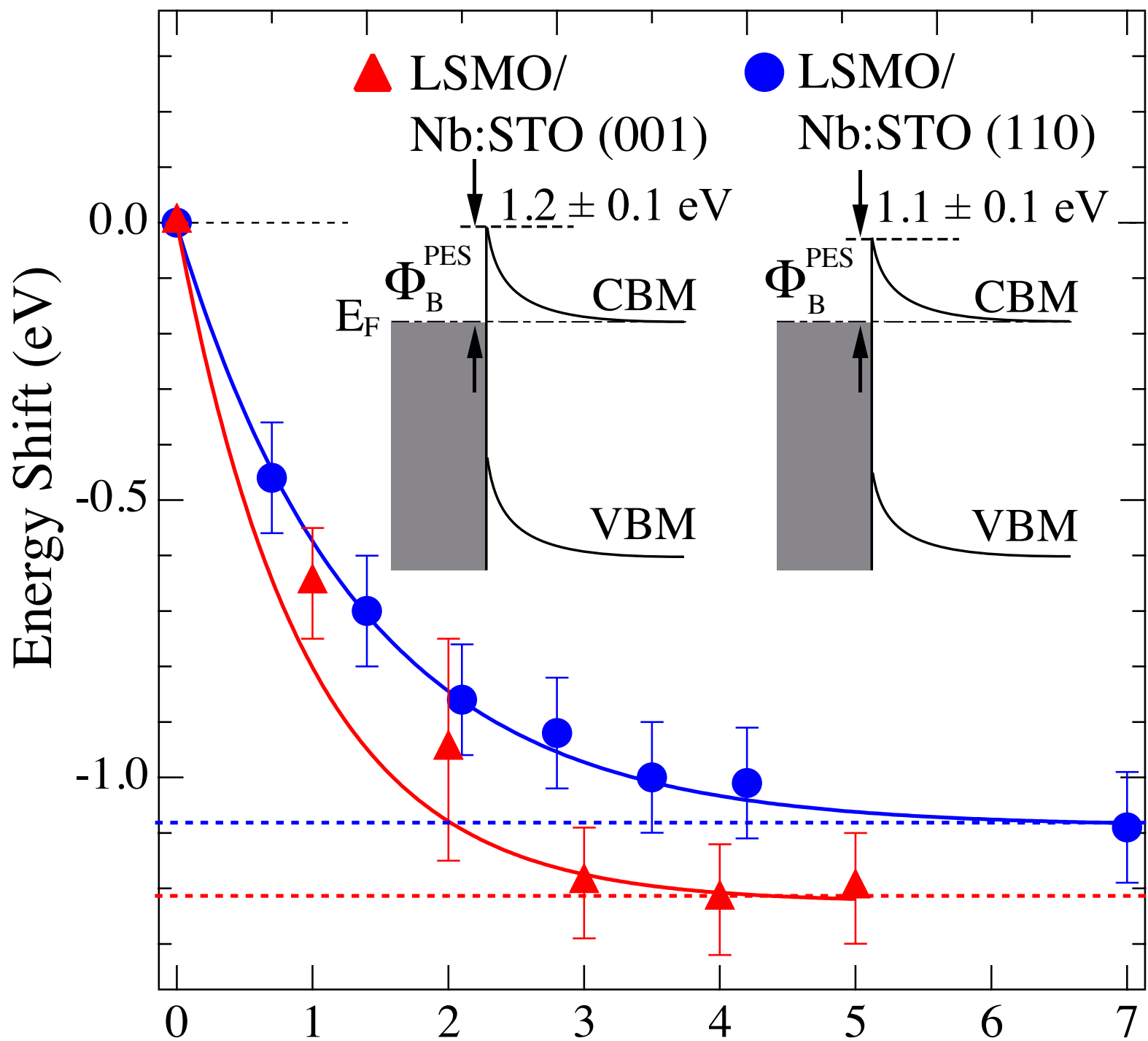